\documentclass[aps,prl,amsmath,amssymb,preprint,showpacs,a4paper,superscriptaddress]{revtex4-1}

\usepackage{graphicx}
\usepackage{dcolumn}
\usepackage{bm}
\usepackage{xcolor}
\usepackage[english]{babel}

\begin{document}	

\title{Abiotic molecular oxygen production - ionic pathway from sulphur dioxide}

\author{Måns Wallner}
\affiliation{Department of Physics, University of Gothenburg, Origov\"agen 6B, 412 58 Gothenburg, Sweden}

\author{Mahmoud Jarraya}
\affiliation{Université de Tunis El Manar, Faculté des Sciences de Tunis, Laboratoire de Spectroscopie Atomique, Moléculaire et Applications – LSAMA, 2092, Tunis, Tunisia}
\affiliation{Université Gustave Eiffel, 
COSYS/LISIS, 5 Bd Descartes, 77454 Champs sur Marne, France}

\author{Saida Ben Yaghlane}
\affiliation{Université de Tunis El Manar, Faculté des Sciences de Tunis, Laboratoire de Spectroscopie Atomique, Moléculaire et Applications – LSAMA, 2092, Tunis, Tunisia}

\author{Emelie Olsson}
\affiliation{Department of Physics, University of Gothenburg, Origov\"agen 6B, 412 58 Gothenburg, Sweden}

\author{Veronica Ideböhn}
\affiliation{Department of Physics, University of Gothenburg, Origov\"agen 6B, 412 58 Gothenburg, Sweden}

\author{Richard J. Squibb}
\affiliation{Department of Physics, University of Gothenburg, Origov\"agen 6B, 412 58 Gothenburg, Sweden}

\author{Gunnar Nyman}
\affiliation{Department of Chemistry and Molecular Biology, University of Gothenburg, 
Kemig\aa rden 4, 412 96 Gothenburg, Sweden} 

\author{John H.D. Eland}
\affiliation{Department of Chemistry, Physical and Theoretical Chemistry Laboratory, Oxford University, South Parks Road, Oxford OX1 3QZ, United Kingdom} 

\author{Raimund Feifel}
\email{raimund.feifel@physics.gu.se}
\affiliation{Department of Physics, University of Gothenburg, Origov\"agen 6B, 412 58 Gothenburg, Sweden} 

\author{Majdi Hochlaf}
\email{majdi.hochlaf@univ-eiffel.fr}
\affiliation{Université Gustave Eiffel, 
COSYS/LISIS, 5 Bd Descartes, 77454 Champs sur Marne, France}

\date[]{}

\begin{abstract}
Molecular oxygen, O$_2$, is vital to life on Earth and possibly on other planets. Although the biogenic processes leading to its accumulation in Earth's atmosphere are well understood, its abiotic origin is still not fully established. Here, we report combined experimental and theoretical evidence for electronic-state-selective production of O$_2$ from SO$_2$, a major chemical constituent of many planetary atmospheres and one which played an important part on Earth in the Great Oxidation event. The O$_2$ production involves dissociative double ionisation of SO$_2$ leading to efficient formation of the O$_2^+$ ion which can be converted to abiotic O$_2$ by electron neutralisation. We suggest that this formation process may contribute significantly to the abundance of O$_2$ and related ions in planetary atmospheres, especially in those where CO$_2$, which can lead to O$_2$ production by different mechanisms, is not the dominant component.

\end{abstract}

\maketitle	


Molecular oxygen, O$_2$, is closely connected to life on Earth and has a decisive role in the oxidation state of the other elements. Before becoming stabilised, the O$_2$ concentration in the atmosphere increased strongly during the Great Oxidation Event which occurred $\sim$2.4 billion years ago \cite{Holland_06}. Although the biogenic processes leading to its accumulation in the atmosphere are well understood, its abiotic origin is still not fully established. In primitive Earth, the commonly admitted models assume that O$_2$ is formed through the three-body recombination reaction O + O + M $\to$ O$_2$ + M \cite{Holland_06,Kasting_84,Kasting_81,Kasting_79}. In contrast, the pioneering work of Ng and co-workers from 2014 \cite{Lu_14} showed that state selective VUV photodissociation of CO$_2$ can lead to O$_2$ with 5\% efficiency. Two years later, Tian and co-workers \cite{Wang_16} provided experimental evidence for a channel of dissociative electron attachment to CO$_2$ that produces O$_2$ + C$^-$. Also, preliminary work by Larimian et al. \cite{Larimian_17} suggests as possible reaction path CO$_2^{2+}$ $\to$ O$_2^+$ + C$^+$. Since CO$_2^{2+}$ dications were detected in the outer atmospheres of those Solar System bodies \cite{Gu_20,Beth_20}, the high O$_2^+$ concentration in the ionospheres of both Venus and Mars may have some contribution from CO$_2^{2+}$ dissociation. However, these findings cannot explain the high abundance of O$_2$ and related ions in other planetary atmospheres where CO$_2$ is not a dominant component. As one example, O$_2^+$ is the most abundant molecular ion in the ionosphere of Venus. This is surprising, since in this atmosphere there is relatively little ($<0.2\%$) O$_2$ to ionise \cite{Hanson_77,Drake_96}. An alternative explanation has been proposed 
based on the fast exothermic reaction O$^+$ + CO$_2$ $\to$ O$_2^+$ + CO, where the O$^+$ ions are released from dissociative ionisation of CO$_2$. However, this mechanism cannot play a major role either because of the low CO$_2$ concentration. 

Sulphur dioxide, SO$_2$, is an important chemical compound in the context of the Great Oxidation Event \cite{Holland_06} where it strongly influenced the Earth's first sulphur cycle through its participation in gas phase reactions resulting in fixing the oxidation state of sulphur \cite{Farquhar_00}. It is known that it drives the geochemical cycle of sulphur in the Earth's, Mars', Venus', Io's  and exoplanet terrestrial atmospheres \cite{Broadfoot_79,Zhang_10,Pearl_79,Lellouch_07}, leading, for instance, to high SO$_2$ concentrations in the atmospheres of Venus \cite{Stewart_79,Zhang_10} and Io \cite{Russell_00} mainly due to outgassing from volcanism. In fact, more than 90\% of Io's atmosphere is composed of SO$_2$ \cite{Na_90,Barker_79,Lellouch_07}. 
On Earth, it has both anthropogenic and natural origins, and it is involved in the formation of environmental pollutants (e.g. sulphuric acid or sulphuric acid aerosols \cite{Hu_13}), contributing to the related deleterious effects such as acid rains and smogs. 

Apart from that, SO$_2$ may have played a role in the emergence of life on Earth either directly or indirectly by contributing to O$_2$ formation in form of abiotic processes. Indeed, it was suggested recently that SO$_2$ clouds escaping from erupting volcanoes may have kickstarted chemical processes that led to the emergence of life on Earth more than four billion years ago \cite{Ranjan_18}. 
While its role in the formation of other sulphur containing compounds is relatively well understood, the involvement of SO$_2$ in the production of other major components such as O$_2$ and O$_2^+$ in the atmospheres of Earth, Venus or Io is not. For neutral SO$_2$, ultrafast photodissociation dynamics studies induced by intense ultrashort laser pulses in a pump-probe scheme evoked the idea that an intermediate fast rotating O$_2$ molecule might be formed before complete fragmentation \cite{Lin_20}, but no O$_2$ production from SO$_2$ was revealed in these experiments. 

The present work shows that dissociative double ionisation of SO$_2$ produces O$_2^+$  which after neutralisation may form abiotic O$_2$. This can be regarded as the first experimental evidence of an ionic pathway for the formation of O$_2$ in abiotic media. In the present-day atmosphere of Io and the primitive atmospheres of Earth and Venus, ionic pathways from SO$_2$ to O$_2$ are certainly plausible since those media were strongly bombarded by ionising radiation where single, double and inner shell ionisations are very likely to occur. In the context of the formation of O$_2^+$ from SO$_2^{2+}$ our calculations show that the less stable non-linear SO$_2^{2+}$ and especially OOS$^{2+}$ isomers  play crucial roles. The OSO$^{2+}$ isomer is linear with two far O atoms whereas the two O atoms are directly bound to each other in OOS$^{2+}$. Therefore, we mapped the potential energy surfaces of OSO$^{2+}$ and those of OOS$^{2+}$ to give insights into the formation mechanism of O$_2^+$ by state selective double ionisation of SO$_2$ (cf. Figs. S\textcolor{orange}{??} for more details).

\section{Results and Discussion}

SO$_2$ has been investigated experimentally in the present work and in the past \citep{Hochlaf_04} by multi-particle coincidence detection using the TOF-PEPEPIPICO technique where the target species were irradiated by 40.81 eV photons from a pulsed helium discharge lamp. From previous studies (cf. Ref. \cite{Jarray_21} and refs. therein) it is known that after double ionisation of SO$_2$ dissociates in four principal channels:
\begin{align}
\text{SO}_2^{2+} &\to \text{O}^+ + \text{SO}^+ \\
\text{SO}_2^{2+} &\to \text{O}^+ + \text{S}^+ + \text{O} \\
\text{SO}_2^{2+} &\to \text{O}_2^+ + \text{S}^+ \\
\text{SO}_2^{2+} &\to \text{O} + \text{SO}^{2+}.
\end{align}
A fifth channel producing two O$^+$ and neutral S is detectable but of negligible intensity. 

The SO$_2^{2+}$ $\to$ O$_2^+$ + S$^+$ channel, which is of main interest in here, is embedded in a dense manifold of dissociation limits forming SO$^+$ + O$^+$ and SO$^{2+}$ + O products in their electronic ground states. Standard unimolecular reaction rate theory, where all open channels compete, predicts that products at the lowest dissociation limit (i.e. SO$^+$ (X$^2\Pi$) + O$^+$ ($^4$S)) should be thermodynamically favoured over fragments from less stable channels (cf. Table S3 of Ref. \cite{Jarray_21}). In contrast, we demonstrate here that the higher energy product O$_2^+$ is formed by state selective dissociation of SO$_2^{2+}$; the O$_2^+$  can then form O$_2$ by electron recombination. Besides the importance of these findings for the abiotic generation of O$_2$, a new mechanism is proposed that may apply widely in dissociation of doubly and multiply ionised molecules.

\begin{figure}[h]
\includegraphics[width = 1\textwidth]{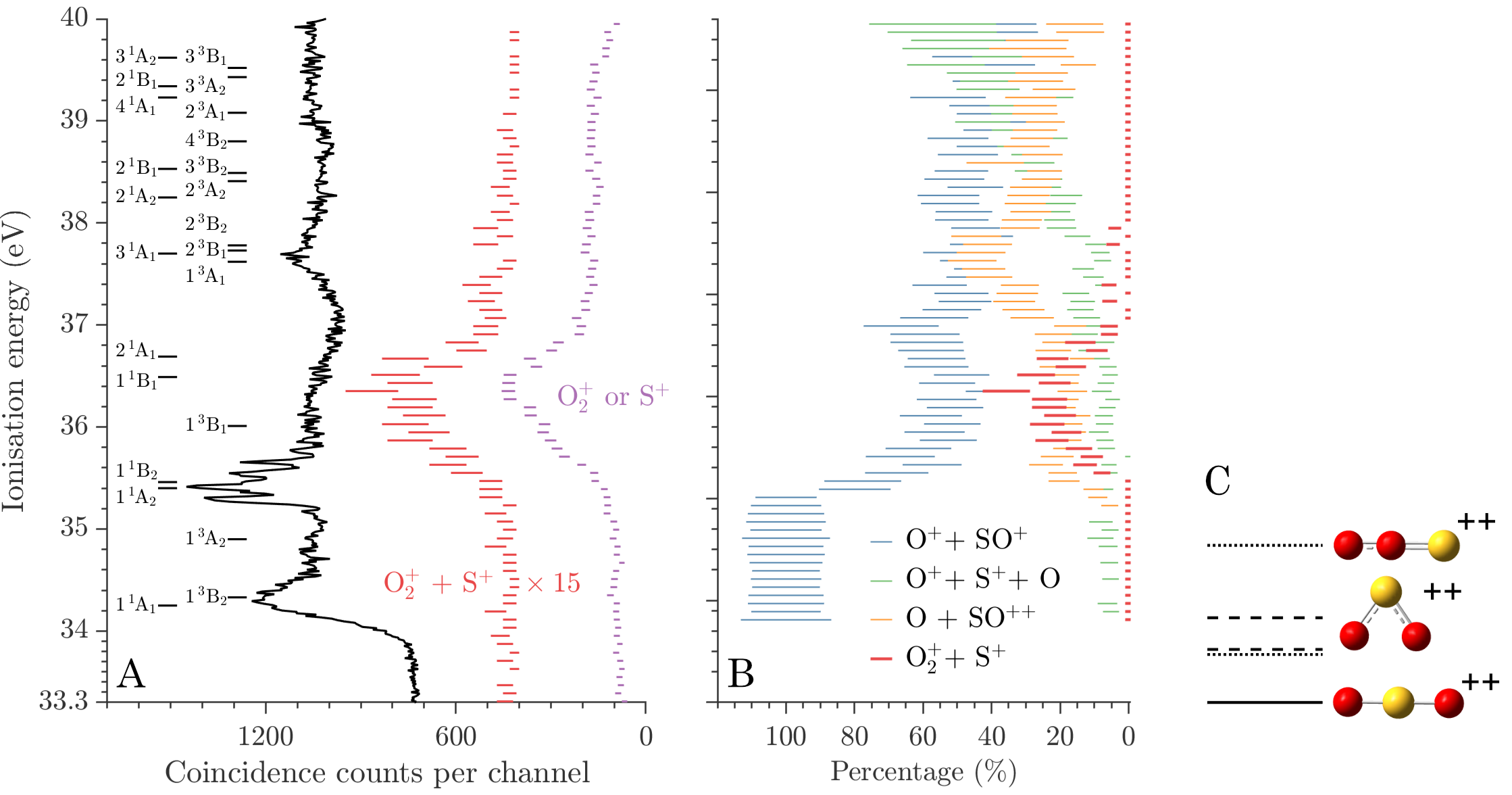}
\caption{Panel A shows electron pairs measured in coincidence with two ions in red (from fourfold events) and one ion in purple (from threefold events) upon photoionisation of SO$_2$ at $40.81$ eV photon energy. For comparison, the black curve is a higher resolution electron pair only spectrum of the total double ionisation at the same photon energy previously discussed in Ref. \cite{Eland_03}. The bar combs mark the vertical ionisation energies computed at the MRCI/aug-cc-pV(Q+d)Z level of theory at the neutral SO$_2$ [X $^1$A$_1$] ground state equilibrium geometry, i.e. at an O--S--O angle of $120^\circ$ and $R=2.7$ Bohr. Panel B shows a breakdown diagram of the major detectable decay channels of doubly ionised SO$_2$. Panel C indicates schematically the stable isomers of SO$_2^{2+}$, which are accessible in the energy region where O$_2^+$ + S$^+$ is detectable.}
\label{fig:exp}
\end{figure}

Panel A of Fig. \ref{fig:exp} shows an electron pair spectrum extracted in correlation with either O$_2^+$ or S$^+$ ions (purple curve) from three-fold coincidence events, as well as an electron pair spectrum extracted in correlation with both the O$_2^+$ and S$^+$ ions (red curve) from fourfold events. These two spectra were obtained under identical conditions. 
For comparison, a higher resolution electron-pairs-only spectrum which reflects the total double ionisation at the same photon energy, previously discussed in Ref. \cite{Eland_03}, is shown in black. The ion fragmentation curves are displaced horizontally on a common scale (with fixed offsets) and show true relative intensities, while the overall spectrum included for comparison has an intensity scale adjusted for display. The horizontal bar combs mark MRCI/aug-cc-pV(Q+d)Z computed vertical double ionisation energies of SO$_2$ quoted for C$_{2v}$ symmetry. 

Panel B of Fig. \ref{fig:exp} presents an estimate of the branching ratios of SO$_2$ for the experimentally detectable dissociative double ionisation channels at 40.81 eV photon energy. For the channels involving two ionic fragments fourfold coincidences are used, and for the channel containing O + SO$^{2+}$ threefold coincidences are used. The red curve shows the channel leading to O$_2^+$ + S$^+$ and corresponds to the red curve in panel A. 

Panel C of Fig. \ref{fig:exp} indicates schematically the energy levels for the isomers of SO$_2^{2+}$ where the linear O-S-O$^{2+}$ is represented by a solid line, the bent O-S-O$^{2+}$ is represented by dashed lines, and the linear O-O-S$^{2+}$ is represented by a dotted line. Because all the isomers appear well below the appearance energy of O$_2^+$ + S$^+$ they are all possible candidates from which the dissociation can occur. OSO$^{2+}$ was already identified in Ref. \citep{Hochlaf_04}, whereas bent OSO$^{2+}$ and OOS$^{2+}$ are identified here for the first time, although a neutral OOS in solid argon was already characterised by Y.-P. Lee and coworkers by IR spectroscopy \cite{chen_96}. In the context of the formation of O$_2^+$ from SO$_2^{2+}$, the less stable bent SO$_2^{2+}$ and especially OOS$^{2+}$ isomers should play crucial roles. Indeed, OSO$^{2+}$ is linear with two far O atoms whereas the two O atoms are bound in OOS$^{2+}$. Therefore, we mapped the potentials energy surfaces of OSO$^{2+}$ and those of OOS$^{2+}$ to give insights into the formation mechanisms of O$_2^+$ by state selective double ionisation of SO$_2$ (cf. Figs. S\textcolor{orange}{??} for more details). 

Insights into the fragmentation mechanisms can be obtained from the kinetic energy releases (KERs). The magnitudes of the KERs in the different dissociation channels can be extracted from the width and shape of the ion TOF peaks, and for the charge-separating channels of doubly-ionised SO$_2$ this has been done several times before \cite{Curtis_85,Masouka_01,Eland_86,Eland_87,Hsieh_95,Hsieh_97} with generally concordant results, but without initial state selection. The most detailed measurements, made possible by use of a position-sensitive ion detector, gave the full KER distributions for 40.81 eV photoionisation \cite{Hsieh_97}. For the two-body dissociation producing O$_2^+$ + S$^+$, the distribution is quite narrow (FWHM $\approx$ 2 eV) and centred at about 5 eV. Field and Eland \cite{Field_99} determined energy releases as a function of the maximum (double) ionisation energy for all four major dissociation channels of SO$_2^{2+}$ using single-electron-ion(-ion) coincidences. For the main two-body charge separation, the energy release was found to vary slightly, between $4.3\pm 0.5$ to $5.2\pm 0.5$ eV over the range of ionisation energies of 34 to 40 eV. Because we collect and analyse both electrons, we can now determine mean KERs as a function of the actual energies transferred in double ionisation, which is an important aspect of the present work.

As can be seen from Fig. \ref{fig:exp}, the channel leading to O$_2^+$ + S$^+$ arises specifically from states in the energy range between 35.5 and 36.5 eV where the $^3$B$_1$, $^1$B$_1$ and $^1$B$_2$ states are found to be located \cite{Hochlaf_04}. Its peak intensity occurs at about 36.2 eV ionisation energy with an onset at about 35.3 eV, energies that are well above the thermodynamic threshold of 28.4 eV for the ground state O$_2^+$ + S$^+$ products. There is a broad feature in the overall double ionisation spectrum at exactly the same energy.  If the mean KER of 6 eV \cite{Curtis_85} or 5.6 eV \cite{Hsieh_97} is added to the ground state product energy, the resulting calculated threshold is 34.3 or 33.9 eV, i.e. below the observed onset. This suggests that the fragments are formed with substantial internal energy.  
In fact, the KER appears from our data to be slightly lower than the values reported by Curtis and Eland ($6\pm 1$ eV) \cite{Curtis_85} and Hsieh and Eland (peak at 5.4 eV) \cite{Hsieh_97}, where the values known in the literature reflect kinetic energies from the whole span of ionisation energies accessed at 40.8 eV. The value we consider most reliable is $4.7\pm 0.3$ eV for the ionisation energy range of 36.3 to 36.8 eV where these products are formed most abundantly. When subtracted from the mean ionisation energy of 36.6 eV, this places the energy of the products at 31.9 eV, well above the thermodynamic limit of 28.4 eV for formation of ground-state products.

\begin{figure}[b]
\includegraphics[width = 1\textwidth]{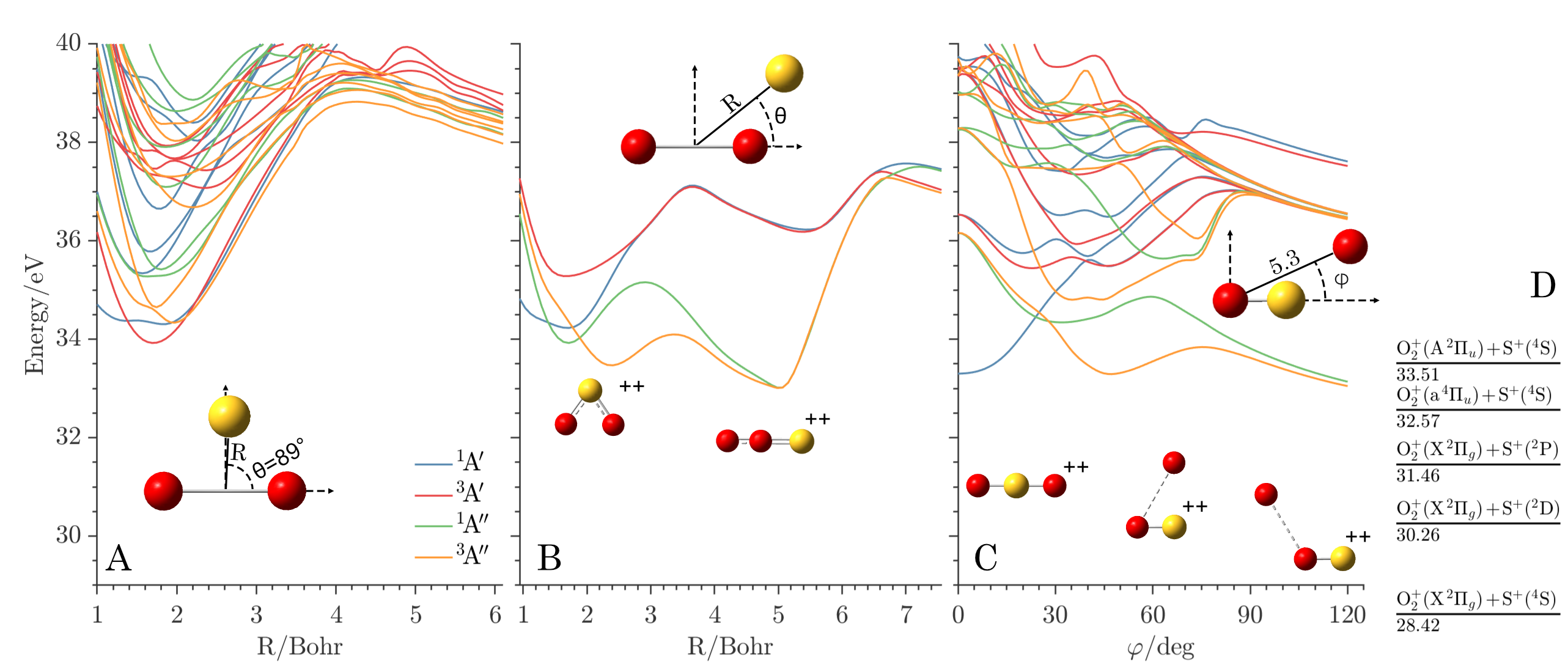}
\caption{Panel A shows a PES cut at the neutral SO$_2$ [X $^1$A$_1$] ground state equilibrium geometry. Panel B shows the corresponding MEP for two stable geometries, bent SO$_2^{2+}$ and O-O-S$^{2+}$. Panel C shows PES with fixed $r$ and varying angle $\varphi$ from 0$^\circ$ to 180$^\circ$. Panel D indicates the thermodynamic thresholds of O$_2^+$ + S$^+$. The reference energy is that of SO$_2$ [X $^1$A$_1$] in its vibrationless ground state. \textcolor{red}{Other cuts are given in Fig. S?? of the SI}}
\label{fig:isomer}
\end{figure}

To further examine the dissociation producing O$_2^+$ + S$^+$ we have performed \textit{ab initio} computations of the thermodynamic thresholds for the SO$_2$ fragments and the potential energy surfaces (PESs) of the singlet and triplet electronic states located in the 32-42 eV energy range above the neutral SO$_2$ [X $^1$A$_1$] ground state. The PES computations were done at the CASSCF/MRCI/aug-cc-pV(Q+d)Z level of theory. The thermodynamic threshold computations were done at the (R)CCSD(T)/CBS level of theory. We note that while the thresholds for singly charged and neutral species are known from thermochemical heats of formation, those of excited and doubly charged species must be estimated theoretically.

Panels A, B and C of Fig. \ref{fig:isomer} show several sets of PES cuts, and panel D presents the thermodynamic thresholds for O$_2^+$ + S$^+$. The PES in panel A and B are computed in Jacobi coordinates where the position of the sulphur atom is varied along the coordinates $\text{R}\in [0.5,\; 5] $ (\AA) and $\theta \in [0,\; 180]$ ($^\circ$) relative to the centre of mass of O$_2$. The oxygen atoms are separated by the equilibrium distance of the neutral SO$_2$ [X $^1$A$_1$] ground state. Panel C shows PES cuts computed in the coordinate system where one oxygen atom is moved relative the other oxygen atom by varying the angle $\varphi$ with the bond distance $\text{R}$ fixed to 5.3 Bohr from the equilibrium of the SO$_2^{2+}$ [X $^1\Sigma^+_g$] ground state.

The curves in panel A of Fig. \ref{fig:isomer} demonstrate that the potential energy barrier is much too high ($\sim 39$ eV), suggesting that the sulphur atom does not leave the system in this way. The vast number of states presented in this figure motivates a minimal energy path (MEP) analysis. Thus, in panel B of Fig. \ref{fig:isomer} the MEP of the $^1\hspace{-1pt}$A$'$, $^1\hspace{-1pt}$A$''$, $^3\hspace{-1pt}$A$'$ and $^3\hspace{-1pt}$A$''$ PES are shown, where the angle $\theta$ with the minimal energy is selected for each $\text{R}$. The MEP analysis suggests a roaming pathway where SO$_2^{2+}$ may convert from a bent O-S-O configuration to a linear O-O-S isomer by crossing a potential barrier at about 34.5 eV. Panel C further demonstrates this roaming mechanism by starting from a linear O-S-O$^{++}$, crossing the barrier at 34.4 eV by spin-orbit conversion from the $^1\hspace{-1pt}$A$'$ to the $^3\hspace{-1pt}$A$''$ state.

\begin{figure}[ht]
\includegraphics[width=1\textwidth]{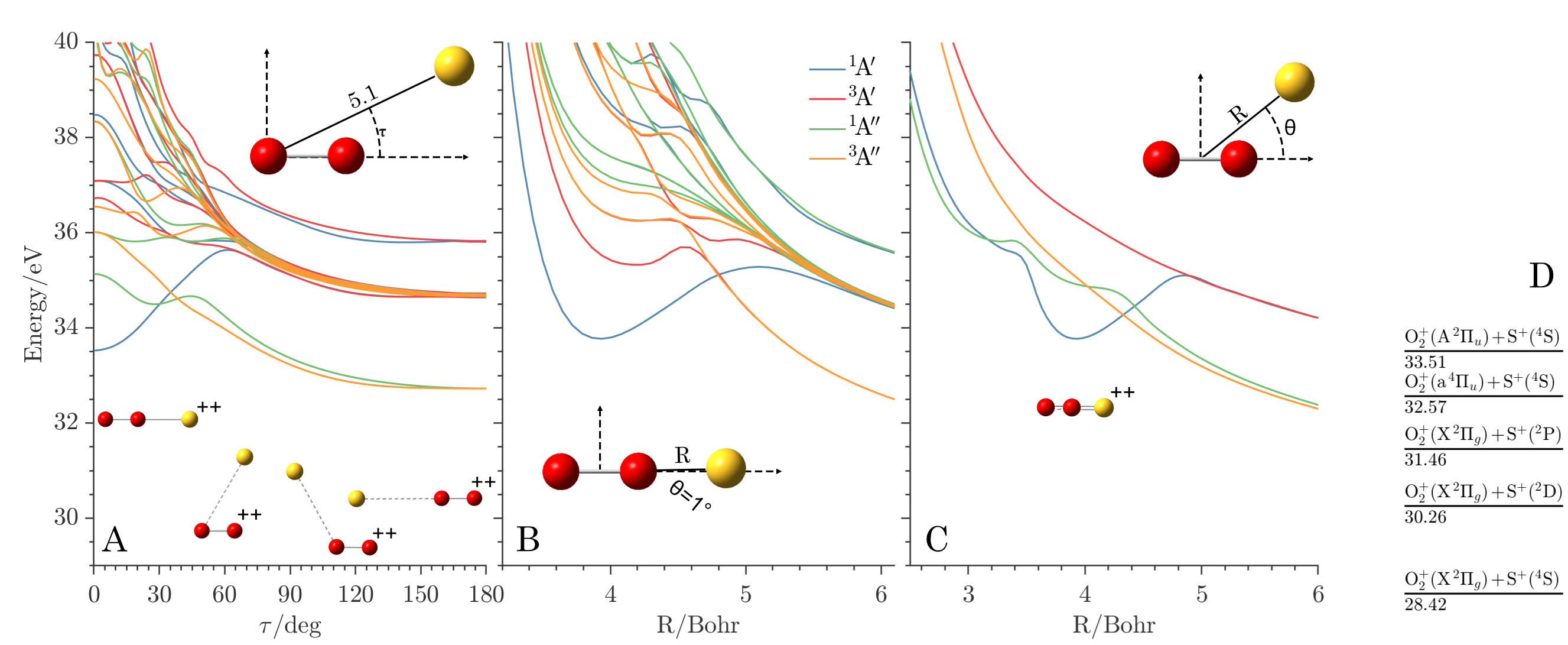}
\caption{Panel A shows PES with fixed $R$ and varying angle $\varphi$ from 0$^\circ$ to 120$^\circ$. Panel B shows a PES cut at the O-O-S quasi-linear geometry with an O-O distance at the neutral O$_2$ [X $^2\Pi_g$] ground state equilibrium geometry. Panel C shows MEPs with a stable $^1$A$'$ state in quasi-linear configuration. In panel D the thermodynamic thresholds of O$_2^+$ + S$^+$ are indicated. The reference energy is that of SO$_2$ [X $^1$A$_1$] in its vibrationless ground state.}
\label{fig:dis}
\end{figure}

Panels A, B and C of. Fig. \ref{fig:dis} show another three sets of PES cuts while panel D shows the associated thermodynamic thresholds of O$_2^+$ + S$^+$. Here, panel A shows PES cuts computed in the coordinate system where one oxygen atom is moved relative the other oxygen atom by varying the angle $\tau$ with $\text{R}$ fixed to 5.1 Bohr from the equilibrium of the OOS$^{++}$ [X $^1\Sigma^+$] ground state. The PES in panels B and C are computed in Jacobi coordinates where the position of the sulphur atom is varied along the coordinates $\text{R}\in [0.5,\; 5]$ (\AA) and $\theta \in [0,\; 180]^\circ$ relative to the centre of mass of O$_2$. The oxygen atoms are separated by the equilibrium distance of the neutral O$_2$ [X $^2\Pi_g$] ground state.

Panel A of Fig. \ref{fig:dis} illustrates how the formation of O$_2^+$ can go through the formation of the OOS$^{2+}$ isomer i.e. after O roaming, since the two O atoms are far away in OSO$^{2+}$. Panels B and C show PES cuts and the MEPs, respectively, and present a decay pathway from the linear configuration O-O-S where the sulphur atom leaves the O$_2$ molecule by crossing the potential energy barrier in a transition from the 1A$'$ to the 3A$''$ state at around 34.5 eV, which is where in fact the experimental data starts showing a signal. The KER obtained by going to the lowest final state is about 6 eV, which is roughly 1.5 eV more than from the present experiment. The excess energy may suggest that some of the energy is converted to vibrational or rotational energy for the O$_2$ moiety. Another explanation could be that the sulphur atom does not end up in the ground state but instead the S [$^2$D] state, which is about 2 eV above the ground state.

\section{Conclusions}
Here we report the first evidence for the production of O$_2$ from SO$_2$. We have established that O$_2^+$ is produced efficiently by dissociative double ionisation of SO$_2$ induced by absorption of 30.4 nm radiation on the HeII$\alpha$ line, an intense component of the solar spectrum and of many stellar spectra. We believe that the same electronic states of nascent SO$_2^{2+}$ that lead to O$_2^+$ formation will be created by all forms of double ionisation, whether by high energy photon or charged particle impact.  The process may therefore be significant in the abiotic formation of molecular oxygen in planetary atmospheres rich in SO$_2$.  The mechanism of the dissociation reaction revealed by our calculations involves an intermediate roaming isomerization; similar mechanisms may be relevant to dicationic dissociations in a range of other molecules.



\nocite{*}
\bibliography{mybib}

\section{Methods}

\subsection{Experimental method}
Multi-electron-ion coincidence experiments were carried out at the photon energy of 40.81 eV which is well above the SO$_2$ adiabatic double ionisation threshold of 33.5 eV \cite{Hochlaf_04}. Our experimental setup allows for simultaneous detection of all ionic species in form of m/q measurements, and all emitted electrons with kinetic energy information. In brief, an effusive jet of target gas is let into the light-matter interaction chamber by a hollow needle where it is intersected by the photon beam for ionisation. A small electric field is applied across the source region to ensure collecting all emitted electrons including those with near-zero kinetic energy. The electrons are guided by means of a strong permanent magnet, which gives rise to a divergent magnetic field capturing electrons in an almost 4$\pi$ solid angle and directing them into a flight tube surrounded by a homogeneous solenoid. This 2.2 m long tube is terminated at the other end by a multi-channel-plate detector. After about 150 ns, a strong electric field is applied to extract the ions in the opposite direction. The ions are guided by a two-field configuration adjusted to time-focusing conditions to achieve optimal mass resolution. At the time of the experiment the collection efficiency was about 30\% for the electrons and about 10\% for the ions (at m/q = 100). The electron resolution was $E/\Delta E = 20$ and the ion mass resolution ranged from 30 to 100 (FWHM).

\subsection{Electronic structure calculations}
The electronic structure calculations were performed using the complete active space self-consistent field (CASSCF) and multi-reference configuration interaction (MRCI) methods, as well as the couple cluster with perturbative triple excitation approach (CCSD(T)) as available in the MOLPRO program suite \cite{MOLPRO_brief}. The Dunnings basis set aug-cc-pV(Q+d)Z \cite{Dunning_89,Kendall_92,Woon_93} has been used for the potential energy surface computations. For the CCSD(T) computations both the aug-cc-pV(Q+d)Z and aug-cc-pV(5+d)Z basis sets were used in parallel for all computations and the output parameters were subsequently fitted to $E_{X} = E_{CBS} + A/X^3$ \cite{Helgaker_97,Puzzarini_09} for complete basis set accuracy. 

The potential energy surfaces were mapped using the CASSCF method followed by the internally contracted MRCI method in C$_s$ symmetry. The active space comprised all configurations, and for the MRCI computations all configurations with coefficients larger than 0.005 in the CI expansion of the CASSCF reference wave function were used. In the first set of computations, 6 states were computed each in $^1$A$'$, $^3$A$'$, $^1$A$''$ and $^3$A$''$ and mapped at 420 geometry points for two different base geometries. The two geometries, shown in Fig. \ref{fig:isomer}B) and in Fig. \ref{fig:dis}C), comprise the O-O distances fixed according to the equilibrium geometry of SO$_2^{++}$ [X $^1$A$_1$] and O$_2^{+}$ [X $^2\Pi_g$], respectively. The second set of computations, shown in Fig. \ref{fig:isomer}C) and in Fig. \ref{fig:dis}A), comprise the starting geometry at the equilibrium of SO$_2^{++}$ [X $^1\Sigma^+_g$] and OOS$^{++}$ [X $^1\Sigma^+$], respectively. In these plots, the oxygen atom and the sulphur atom, respectively, are roaming around the remaining atom by varying the angle, keeping the distance to the initial furthermost atom fixed.

\section{Data availability}
The datasets generated during and/or analysed during the current study are available from the corresponding authors on reasonable request.

\section{Acknowledgements}

We thank the Swedish Research Council and the Knut and Alice Wallenberg Foundation for financial support. This work was carried out while M.H. was Waernska Guest Professor at the University of Gothenburg. The computations involved the Swedish National Infrastructure for Computing (SNIC) at the Chalmers Centre for Computational Science and Engineering (C3SE) partially funded by the Swedish Research Council through grant no. 2018-05973. M.W. thanks Doc. Vitali Zhaunerchyk for an interesting discussion. 


\section{Author contributions}
J.H.D.E. and R.F. conducted the experimental research. J.H.D.E. and M.W. performed the data analysis. M.W., M.J., S.B.Y. and M.H. carried out the theoretical calculations. All authors discussed the results and contributed to the writing of the manuscript at several instances.

\section{Competing interests}
The authors declare no competing interests.

\end{document}